\documentclass[prd,superscriptaddress,preprintnumbers,amsmath,amssymb,nofootinbib]{revtex4}
\usepackage{color}
\usepackage{ulem}
\usepackage{graphicx}
\begin{document}

\title{Non-Perturbative Yang-Mills Condensate as Dark Energy}

\author{Pietro Don\`a}
\affiliation{Department of Physics \& Center for Field Theory and Particle Physics, Fudan University, 200433 Shanghai, China}
\author{Antonino Marcian\`o}
\email[]{marciano@fudan.edu.cn}
\affiliation{Department of Physics \& Center for Field Theory and Particle Physics, Fudan University, 200433 Shanghai, China}

\author{Yang Zhang}
\affiliation{Key Laboratory for Researches in Galaxies and Cosmology, Department of Astronomy, University of Science and Technology of China, Hefei, Anhui, 230026, China}

\author{Claudia Antolini}
\affiliation{Department of Physics \& Center for Field Theory and Particle Physics, Fudan University, 200433 Shanghai, China}

\date{\today}

\begin{abstract}
\noindent 
Models based on Yang-Mills condensate (YMC) have been advocated in the literature and claimed to be successful candidates to explain dark energy. Several instantiations of this simple idea have been considered, the most promising of which are reviewed here. Nevertheless, results previously attained heavily relied on the perturbative approach to the analysis of the effective Yang-Mills action, which is only adequate in the asymptotically-free limit, and were extended into a regime, the infrared limit, in which confinement is expected.
We show that if a minimum of the effective Lagrangian in $\theta \!=\! - F_{\, \, \mu \nu}^a \, F^{a \mu \nu}/2$ exists, a YMC forms that drives the Universe toward an accelerated de Sitter phase. The details of the models depend weakly on the specific form of the effective Yang-Mills Lagrangian. Using non-perturbative techniques mutated from the functional renormalization group procedure, we finally show that the minimum in $\theta$ of the effective Lagrangian exists, thus YMC can actually take place. The non-perturbative model has properties similar to the ones of the perturbative model. In the early stage of the universe, the YMC equation of state has an evolution that resembles the radiation component, i.e. $w_y \rightarrow 1/3$. However, in the late stage, $w_y$ naturally runs to the critical state with $w_y =-1$, and the universe transits from matter-dominated into dark energy dominated stage only recently, at a redshift the value of which depends on the initial conditions that are chosen while solving the dynamical system. 
\end{abstract}

\maketitle

\section{Introduction}

\noindent
Observational data collected over the last two decades from Supernovae Type I a (SN Ia) confirmed that the Universe is undergoing an accelerated phase of expansion. The first evidences for such a behavior have been discovered by two independent collaborations, and reported in Ref.~\cite{Riess:1998cb}, by the High-redshift Supernova Search Team, and in Ref.~\cite{Perlmutter:1998np}, by Supernova Cosmology Project Team. Analyses from combined  SN Ia dataset (see {\it e.g.} the one reported in Ref.~\cite{Kowalski:2008ez}), cosmic microwave background radiation (CMB), for which we refer for instance to Refs.~\cite{Spergel:2003cb, Graham:2005xx, Spergel:2006hy, Komatsu:2008hk}, and from large scale structure \cite{Cole:2005sx,Tegmark:2006az}, have further provided consolidated evidence for the current acceleration of the Universe. The source of late-time cosmic acceleration has been dubbed as ``Dark Energy'' (DE), an exhaustive theoretical characterization of which is still lacking. There have been many attempts so far to individuate the origin of DE, but a consensus in the literature has not been reached yet. 

A review of the models so far advocated within the literature of DE is beyond the purpose of this study, and we prefer to address reader to a sizable and rich literature that exists on this subject (see {\it e.g.} Refs.~\cite{Capozziello:2003tk, Carroll:2003wy, AT}). 
In what follows, we will focus on a rather simple idea, which address the problem of DE from the perspective of condensation of Yang-Mills fields. Cosmic acceleration as a source of cosmological inflation has been first proposed by Yang Zhang in Ref.~\cite{YZ1}, and then further developed by the same author and collaborators in the framework of current cosmic  acceleration, as a source of DE, in Refs.~\cite{YZ2} and \cite{YZ3}, respectively in the perturbative two-loops and three-loops analysis of the effective action of Yang-Mills theory. 
To be more specific, in Ref.~\cite{YZ1} the author has considered a Yang-Mills gauge boson condensate as described by the renormalization-group-improvement (RGI) action within a homogenous and isotropic FLRW background. Following Refs.~\cite{YZr1, YZr2, YZr3}, the action for SU(N) Yang-Mills fields has been reshuffled in terms of an effective running coupling constant $g=g(\tau)$, namely 
\begin{eqnarray}
\label{first}
\mathcal{S}_{\rm YM}= \int d^4x \sqrt{-{\rm det} (g_{\mu \nu})}\, \mathcal{L}_{\rm eff}\,, \qquad 
\mathcal{L}_{\rm eff} = -\frac{1}{4 g^2(\tau)} F^a_{\mu \nu} F^{a \,\mu \nu}\,, \qquad \tau:=\ln \left| -\frac{F^a_{\mu \nu} F^{a \,\mu \nu}}{2\,  \kappa^2}\right| 
\end{eqnarray}
in which $g_{\mu \nu}$ stands for the background metric and $\kappa$ is the square of the renormalization mass-scale. 
From now on, for simplicity of notation we define the contraction of the field-strength tensors 
\begin{eqnarray}
\theta:=  -\frac{1}{2}\, F^a_{\mu \nu} F^{a \,\mu \nu}\,,
\end{eqnarray}
which plays the role of an order-parameter for the YMC, and allows us to write the effective Lagrangian in a more compact way $\tau=\ln |\theta/ \kappa^2|$. A further simplification within the studies so far developed in the literature concerns the use of a SU(N) gauge symmetry group, in which the number of colors $N$ is not fixed {\it a priori}. This choice connects the physics under investigation to the constituents gauge-groups of the standard model of particles.  

\subsection{The perturbative expansion: the one-loop effective action}

\noindent
Moving from the two-loops perturbative (ultraviolet) expansion at large $\tau/\kappa^2$ of the effective coupling constant, which from the analyses reported in Refs.~\cite{YZr1,YZr2} reads
\begin{eqnarray}
g^{-2}(\tau)=b_0 \tau + 2 \frac{b_1}{b_0} \ln \tau\,,
\end{eqnarray}
the analysis within Ref.~\cite{YZ1} has first focused on a one-loop expansion. This latter entails considering the effective action 
\begin{eqnarray}
\mathcal{L}_{\rm eff}= \frac{1}{2} b_0 \, \theta \ln \left| \frac{\theta}{e \kappa^2} \right| \,.
\end{eqnarray}
The constants $b_0$ and $b_1$ are of order one, and depend on the number of colors $N$ of the SU(N) gauge-group. Relying on the form of the effective Lagrangian that has been derived for the asymptotically free regime, it has been shown that when the minimum is attained, namely when $\theta=\kappa^2$, $g^2\simeq b_0$, the energy density becomes 
\begin{eqnarray}
\rho=\frac{b_0}{2}\,(E^2-B^2)\,, \qquad B^2= \frac{1}{2}\, F_{\, \, ij}^a \, F^{a ij }\,, \qquad E^2= - \frac{1}{2}\, F_{\, \, 0i}^a \, F^{a 0i }\,.
\end{eqnarray}
The equation of state then reads like the one of dark energy, namely 
\begin{eqnarray}
p=-\rho\,.
\end{eqnarray}

\subsection{Two-loops and three-loops effective action, and the issue of stability}

\noindent
The issue of proving that the dark energy behavior of YMC is stable with respect to higher order loop-corrections has been addressed in \cite{YZ2} and \cite{YZ3}. In Ref.~\cite{YZ2}, the analysis has resorted to the two-loops effective Lagrangian,  
\begin{eqnarray}
\mathcal{L}_{\rm eff}= \frac{b}{2} \theta \left[ \ln \left| \frac{\theta}{\kappa^2}\right|+ \eta \ln \left| \ln \left| \frac{\theta}{\kappa^2}\right| +\delta \right| \right]\,,
\end{eqnarray}
derived within the asymptotically-free regime and then extended to the infrared confining regime, in the range of values for $\theta$ in which $\mathcal{L}_{\rm eff}$ has a minimum. Again, the coefficients $b$ and $\delta$ depend on the number of color $N$ of the gauge-symmetry group, and specifically read $b=\frac{11 N}{3 (4 \pi)^2}$ and $\eta=2 \frac{b_1}{b^2}$, with $b_1=\frac{17 N^2}{3 (4 \pi)^4}$. The energy density and the pressure from this model are provided by the relations
\begin{eqnarray}
&&\rho=\frac{b}{2} \theta \left[\tau +2 + \eta \left( \ln |\tau +\delta| + \frac{2}{\tau + \delta}  \right) \right] \,, \\
&&p=\frac{b}{6} \theta \left[\tau - 2 + \eta \left( \ln |\tau +\delta| - \frac{2}{\tau + \delta}  \right) \right] \,.
\end{eqnarray}
At high energies, when $\tau>\!\!>1$, the equation of state of the YMC evolves towards the equation of state of radiation
\begin{eqnarray}
w=\frac{p}{\rho} \rightarrow \frac{1}{3}\,.
\end{eqnarray}
Approaching the infrared regime, at energies such that $\tau< -1.8$, the weak energy condition (see {\it e.g.} \cite{YZr6}) is violated, and the equation of state parameter crosses the value $w=-1$. Notice however that, moving from the expression of $\mathcal{L}_{\rm eff}$ determined in the asymptotically-free regime, while reaching the range of values for $\tau$ that ensure a dark energy behavior for the Universe, a divergence is encountered in the effective coupling constant $g^2(\tau)$. Although divergences do not occur in the observable quantities, which are the pressure $p$, the energy density $\rho$, and their ratio $w$, the occurrence of a divergence in $g^2(\tau)$ makes less trivial the act of relying on the extension of this procedure to higher-order loop expansions. Nevertheless, for the purposes of an analysis of the dark energy behavior, it has been shown that all the physical quantities behave like smooth functions in the range of the infrared values $\tau\in (-2.04, 53)$. Furthermore, the stability of the system accounting for interactions with matter and electromagnetic radiation, into which the YMC may decay, has been successfully checked. To achieve this goal, in \cite{YZ2} the authors have considered: the dynamical system provided by the first Friedmann equation in presence of matter with energy density $\rho_{m}$ and (electro-magnetic) radiation with energy density $\rho_r$
\begin{eqnarray} \label{H}
H^2=\frac{8 \pi G}{3} (\rho + \rho_m +\rho_r)\,;
\end{eqnarray}
the equations of motion for the energy densities component of YMC, matter and radiation that arise from conservation of the total energy-momentum tensor and from the decay of the YMC into matter. In comoving coordinates, in which $H=\frac{\dot{a}}{a}$, these latter equations read
\begin{eqnarray} \label{emt}
&&\dot{\rho} + 3 H (\rho+p)=-\Gamma \rho \,, \nonumber \\
&&\dot{\rho}_m + 3 H \rho_m=\Gamma \rho \,, \nonumber \\ 
&& \dot{\rho}_r + 3 H (\rho_r+p_r)=0\,.
\end{eqnarray}
The decay rate $\Gamma$ is a free parameter of the model that enters the definition of the dimensionless dynamical system to be solved, equation (\ref{H}) and equations (\ref{emt}). The specific value of the rescaled parameter $\gamma=\frac{\Gamma}{H}$ then affects the attractor coefficients in the analysis of the stability of the model. 
Consistency with Big Bang Nucleosynthesis (BBN) \cite{YZr5}, with the fractional density observed for dark energy ($\Omega\sim 0.73$) and with the observational constraint on the equivalence between the energy densities of radiation and matter at the redshift of recombination, then set the initial conditions of the dynamical variables. Finally, the model has been shown to posses a dark energy tracking solution, and the dynamical system to have a fixed point, which is stable. For $\gamma_0=0.5$, the parameter of the equation of state takes the asymptotic value $w=-1.14$, provided that in $\mathcal{L}_{\rm eff}$ one sets up $\delta=3$. A different value of this latter parameter, for instance $\delta=7$, would rather entail $w=-1.18$.

\noindent 
A further analysis, developed along the same lines as the one reported above, has been deepened by the same authors in \cite{YZ3}, and the investigation extended to the case of the three-loops effective action. Within this latter work, the effective three-loops coupling constant reads
\begin{eqnarray}
g^2(\tau)=\frac{1}{b} \left[ \frac{1}{\tau} -\eta \frac{\ln|\tau|}{\tau^2} + \eta^2 \frac{\ln^2|\tau|-\ln|\tau| +C}{\tau^3} +  O\left( \frac{1}{\tau^3}\right)\right] \,,
\end{eqnarray}
and the main predictions have been substantially unaffected by the improvement in the perturbative analysis. For different choices of $\gamma_0$, which is the decay rate parameter rescaled by $H$ evaluated at present time, the parameter of the equation of state at present time takes the values $w_0=1.05$ (given that $\gamma_0=0.31$) or $w_0=-1.15$ (given that $\gamma_0=0.67$). 

We emphasize that in both cases summarized above, the two-loops expansion within \cite{YZ2} and the three-loops expansion within \cite{YZ3}, the dark energy behavior arises from the perturbative computation of the effective Lagrangian in the asymptomatically-free limit. The validity of this procedure is then extended to the infra-red regime of the Yang-Mills theory, in order to recover a minimum in $\theta$ for the effective Lagrangian, and derive the equation of state for the Universe, which entails accelerated expansion. But the occurrence of divergences in the effective action may shed some doubts on the validity of the conclusions for the dark energy behavior of the theory. 

Finally, we would like to point out that the gauge interactions taken into account here might not be necessarily considered to be the ones constituting the standard model of particle physics. It is interesting to notice that an extra ``dark sector'' might be advocated to explain the gauge group here involved. Furthermore, it might be tempting to identify the gauge group copies with suitable candidates for dark matter, postulating the existence of ``dark copies'' of fermions colored under the extra ``dark gauge group''. The dark matter sector that is then introduced may eventually be connected to Mirror Standard Model theories \cite{foot1, foot2} discussed in the literature.

\section{Toward a non-perturbative infrared analysis}

\noindent
Within previous studies \cite{YZ1, YZ2, YZ3}, the existence and the stability of a dark energy mechanism based on YMC has been investigated. The analyses have been developed moving from the ultraviolet results recovered in the literature, up to three-loops in the effective action for Yang-Mills SU(N) fields, and results have been then extended in \cite{YZ1, YZ2, YZ3} to the infrared regime, in order to derive a physical characterization of an accelerated expansion of the Universe. An important technical issue is that one that concerns the stability of the result at higher loops expansion, since the appearance of additional term in the effective action might indeed spoil stability, which totally relies on ultraviolet perturbative expansion.

Furthermore, we know that the infrared regime of Yang-Mills SU(N) theories has a completely different behavior than the ultraviolet regime, the latter encoding asymptotic freedom, while the former showing a confinement behavior that depends on the number of gauge colors involved, and even on the number of colored fermions involved in the theory. 

Therefore, the main motivation of this study has been to prove that under mild and general assumptions, which are basically the existence of a minimum in $\theta$ in the non-perturbative effective Lagrangian, a dark energy behavior is recovered. Then by making use of the non-perturbative techniques mutated from the functional renormalization group procedure, which is more adequate to be used in the confining infrared limit of the theory, it is possible to prove that a such a minimum indeed exists. 

At this purpose, we provided the explicit example of the SU(2) Yang-Mills action, deriving the effective Lagrangian for such a particular model and deepening the cases in which interactions with different forms of matter is considered. The procedure, which might not be completely reliable for the precise determination of the coefficients of the effective non-perturbative Lagrangian, is nevertheless enough to ensure that a minimum of $\theta$ exists for $\mathcal{L}_{\rm eff}$, and thus that a YMC works as reliable candidate to explain the origin of Dark Energy. In the following sections, we unfold detailed arguments in support of this thesis.

\subsection{Yang-Mills effective action from a non-perturbative approach} \label{snp}
\noindent 
Within the perturbative YMC model for dark energy that we have reviewed in the previous section, the effective Yang-Mills Lagrangian is that one calculated at one-loop in the seminal work by Savvidy \cite{Savvidy:1977as}, namely
 \begin{eqnarray}\label{l_eff}
  \mathcal{L}_{\rm eff}=\frac{1}{2}b \, \theta \,  \log \left|\frac{\theta}{\kappa^2}\right|,
 \end{eqnarray}
where $b=(11N)/24\pi^2$ for SU(N). Higher-loops correction have been then deployed in order to check whether the substance of the physical results remains unchanged, and how the details may vary. \\

\noindent 
Instead of keep considering higher-loops improvements of \eqref{l_eff}, the main purpose of this work is to start from a general non-perturbative form of the action, {\it i.e.}
\begin{eqnarray}\label{l_eff_np}
 \mathcal{L}_{\rm eff}= \mathcal{W}\left(\theta \right)\,.
\end{eqnarray}
We may then proceed to determine the form of $\mathcal{W}(\theta)$ by stating some general requirements that must be fulfilled in order to obtain a YMC that can work to explain the origin of dark energy. The function $\mathcal{W}$ will be in general equipped also with an energy scale $\kappa$ in analogy with \eqref{first} for dimensional reasons. We notice indeed that $\mathcal{W}$ must be a (not necessarily analytic) function of $\theta$ satisfying the following properties:
\begin{enumerate}
\item
$\mathcal{W}(\theta)$ has a non trivial minimum in $\theta$; 
\item
$\mathcal{W}(\theta)$ possesses a perturbative limit, which reproduces  the one-loop result derived by Savvidy \cite{Savvidy:1977as};
\item
$\mathcal{W}(\theta)$ shows the asymptotic behavior of being at least linear in $\theta$, which in turn is linear to the bare Yang-Mills action. This latter requirement can be formalized as it follows: in the ultraviolet regime $\theta>\!\!> \kappa^2$ it must hold the limit
\begin{eqnarray}
\frac{d \ln \mathcal{W}}{\!\!d \ln \theta}(\theta) \   \longrightarrow{1}   \,.
\end{eqnarray}          
\end{enumerate}

\noindent 
In what follows we will consider for simplicity a pure SU$(2)$ Yang-Mills theory, the gauge field of which is not coupled to any other fundamental matter fields. We want to stress that SU$(2)$ Yang-Mills fields introduced here, as well as SU(N) Yang-Mills fields dealt with in \cite{YZ1,YZ2,YZ3}, despite being suitable to build a model for cosmic dark energy should not be necessarily identified as Standard Model gauge fields. The introduction of additional copy of SU(N) Yang-Mills fields might anyway allow to make contact with some Mirror Standard Model theories \cite{foot1, foot2} that have been advocated in the literature of dark matter. Again, such a link is not necessary for our purposes and will not be analyzed within this investigation of dark energy YMC. Nevertheless it might represent some interesting directions to be followed in forthcoming studies.

\subsection{YMC as cosmological dark energy}
\noindent
In what follows, we discuss the cosmological consequences of the requirements we demanded above for the non-perturbative effective action, and shed light on the behavior of the YMC for the fate of cosmological dark energy. We will assume a flat Friedmann-Lema\^itre-Robertson-Walker (FLRW) universe, the line element of which can be cast respectively in terms of comoving or conformal coordinates as it follows
\begin{eqnarray}\label{FLRW}
 ds^2=dt^2-a^2(t)\delta_{ij}dx^idx^j=a^2(\tau)[d\tau^2-\delta_{ij}dx^idx^j],
\end{eqnarray}
where the cosmological time $t$ and the conformal time $\tau$ are related through $dt=a d\tau$. We will consider the simplest case of a universe filled only with the YMC, and assume it to be minimally coupled to the gravity. Then the effective action reads\footnote{Following the definition in \cite{Misner:1974qy}, we adopt the sign convention $(-,+,+)$ for the metric, the Riemann tensor and Einstein equation respectively.}
 \begin{eqnarray}
 \mathcal{S}=\int \sqrt{-{g}}~\left[-\frac{\mathcal R}{16\pi G}+\mathcal{L}_{\rm eff}\right] ~d^{4}x.
 \label{S}
 \end{eqnarray}
From now on, we will only denote with ${g}$ the determinant of the metric $g_{\mu\nu}$. $\mathcal{R}$ is the scalar Ricci curvature, and $\mathcal{L}_{\rm eff}$ is the effective Lagrangian of the YMC, described by Eq. (\ref{l_eff_np}). Varying $\mathcal{S}$ with respect to the metric $g^{\mu\nu}$, one obtains the Einstein equation $G_{\mu\nu}=8\pi G\, T_{\mu\nu}$, where the energy-momentum tensor of the YMC is given by
 \begin{eqnarray}\label{T_munu}
 T^{\mu\nu}=\sum_{a=1}^{3}~ {}^{(a)}T^{\mu\nu}=\sum_{a=1}^{3}~ g^{\mu\nu}  \mathcal{W}\left(\theta\right) - 2 \dfrac{\partial \mathcal{W}}{\partial \theta} F_a^{\gamma\mu}F^{a}{}_\gamma^{\phantom{\gamma}\nu}.
 \end{eqnarray}
We may set up a gauge that preserves the isotropy and the homogeneity of the FLRW background. We write gauge fields as functions of the cosmological time $t$, and choose their components so to satisfy $A_0=0$ and $A_i^a=\delta_i^aA(t)$. This choice indeed ensures that the total energy-momentum tensor $T_{\mu\nu}$ is homogeneous and isotropic. 
We then introduce the usual definition of the Yang-Mills tensor, cast in terms of the group structure constants $f^{abc}$, namely
\begin{eqnarray}
 F^{a}_{\mu\nu}=\partial_{\mu}A_{\nu}^a-\partial_{\nu}A_{\mu}^a+f^{abc}A_{\mu}^{b}A_{\nu}^{c}\,.
\end{eqnarray}
For the SU$(2)$ gauge-group to which we are specializing our analysis, the structure constants reduce to $f^{abc}=\epsilon^{abc}$. Furthermore, thanks to the gauge fixing we have selected above, and looking at the case of constant electric field (we will assume for simplicity in the following a vanishing magnetic field), we can rewrite the Yang-Mills tensor in the simplified form 
 \begin{eqnarray}
 F^{a\mu}_{~~\nu}=\left(
 \begin{array}{cccc}
      0 & E_1 & E_2 & E_3\\
     -E_1 & 0 & B_3 & -B_2\\
     -E_2 & -B_3 & 0 & B_1\\
     -E_3 & B_2 & -B_1 & 0
 \end{array}
 \right) = \frac{1}{3}\left(
 \begin{array}{cccc}
      0 & E & E & E\\
     -E & 0 & 0 & 0\\
     -E & 0 & 0 & 0\\
     -E & 0 & 0 & 0
 \end{array}
 \right) \,.
 \end{eqnarray}
This allows to express $\theta$ in a very simple form, {\it i.e.} $\theta=\sum_{i=1}^3E_i^2=E^2$, and each component of the energy-momentum tensor then rewrites as 
 \begin{eqnarray}
 ^{(a)}T_{\mu}^{0}&=&-\frac{1}{3}\mathcal{W}(\theta)\delta^{0}_{\mu}+ \frac{2}{3}\mathcal{W}'(\theta)
 E^2\delta^{0}_{\mu} ,\\
 ^{(a)}T_{j}^{i}&=&\frac{1}{3}\mathcal{W}(\theta) \delta^i_j- \frac{2}{3} \mathcal{W}'(\theta) E^2\delta^i_j \delta^a_j.
 \end{eqnarray}
These tensors are not yet isotropic, their values depending on the direction of the color $a$. Nevertheless, the total energy-momentum tensor $T_{\mu\nu}=\sum_{a=1}^3~^{(a)}T_{\mu\nu}$ is isotropic, and 
the corresponding energy density and pressure are given by
\begin{eqnarray} \label{rho_p}
\rho_\text{YMC}=-\mathcal{W}(\theta)+ 2\mathcal{W}'(\theta)\,\theta, \qquad p_{\text{YMC}}=\mathcal{W}(\theta)- \frac{2}{3}\mathcal{W}'(\theta)\,\theta \,.
\end{eqnarray}
Consequently the equation of state (EOS) of the YMC is immediately recovered to be
 \begin{eqnarray}\label{eos}
 w_\text{YMC}\equiv\frac{p_\text{YMC}}{\rho_\text{YMC}}=-\frac{\mathcal{W}-\frac{2}{3}\mathcal{W}' \, \theta}{ \mathcal{W} -2 \mathcal{W}' \, \theta} = -\frac{1-\frac{2}{3}\frac{\mathcal{W}'}{\mathcal{W}} \, \theta}{1 -2 \frac{\mathcal{W}'}{\mathcal{W}}  \, \theta} \, .
 \end{eqnarray}
It is worth discussing the mathematical properties of the EOS $w_\text{YMC}$. On one hand, if we require the Yang-Mills theory to condensate, then the function $\mathcal{W}$ must has a non trivial minimum, as we demanded in Sec. \ref{snp}. But this is equivalent to require that $\mathcal{W}'$ vanish at some point $\theta_0$. In this point the YMC has an EOS of the cosmological constant with $w_\text{YMC}=-1$. Around this critical point the YMC dark energy models can account either for an EOS characterized by $w_\text{YMC}>-1$ or for an EOS characterized by $w_\text{YMC}<-1$, thus encoding phantom matter behavior. 
On the other hand, in the high-energy-scale regime, in which with $\theta\gg \kappa^2$, one would like to recover that the YMC exhibits an EOS of radiation, characterized by $w_\text{YMC}=1/3$, in analogy with the perturbative analysis \cite{YZ1,YZ2,YZ3}. Within the framework of the non-perturbative action \eqref{l_eff_np}, this corresponds to the third requirement listed in Sec. \ref{snp}. The effective action should then scale for $\theta\gg \kappa^2$ at least like the bare Yang-Mills action, {\it i.e.} $d\ln \mathcal{W}(\theta)/d\ln \theta \sim 1$.\\

\noindent
In the following sections, we will show in detail that the YMC evolves from the EOS with $w_\text{YMC}=1/3$ to $w_\text{YMC}=-1$ while the Universe is expanding.

\subsection{A non-interacting YMC model}
\noindent 
The cosmological model we are about to analyze in this section entails three different sources for the energy-momentum tensor: i) the dark energy, the role of which we assume to be played by the YMC; ii) the matter, including both baryons and dark matter, which is dealt with as non-relativistic dust with negligible pressure; iii) the radiation, the component of which consists of photons and possibly other massless particles, such as neutrinos. We will describe the three components in terms of their EOS, without accounting for any microscopic treatment in terms of fundamental matter fields.\\

\noindent
Since from Eq.~(\ref{FLRW}) we assumed {\it ab initio} the universe to be flat, fractional densities will sum up to one, namely $\Omega_\text{YMC}+\Omega_m+\Omega_r=1$. Indeed, fractional energy densities are defined as $\Omega_\text{YMC}\equiv \rho_\text{YMC}/\rho_{tot}$,   $\Omega_{m}\equiv \rho_m/\rho_{tot}$,  $\Omega_{r}\equiv \rho_r/\rho_{tot}$, and the total energy density is given by $\rho_{tot}\equiv\rho_\text{YMC}+\rho_m+\rho_r$.  The overall expansion of the universe is determined by the Friedmann equations:
\begin{eqnarray}\label{friedmann1}
 &&\left(\frac{\dot{a}}{a}\right)^2=\frac{8\pi G}{3}(\rho_\text{YMC}+\rho_m+\rho_r),\\ \label{friedmann2}
 &&\frac{\ddot{a}}{a}=-\frac{4\pi G}{3}(\rho_\text{YMC}+3p_\text{YMC}+\rho_m+\rho_r+3\rho_r),
\end{eqnarray}
where the \textit{dot} denotes the $d/dt$. 

As a first preliminary investigation, we will assume that there is no interaction between the three energy components. The dynamical evolution of these latter is determined by their equations of motion, which in turn follow from imposing the conservation of each  component of the energy-momentum tensor,
 \begin{eqnarray}\label{aa1}
 &&\dot{\rho}_\text{YMC}+3\frac{\dot{a}}{a}(\rho_\text{YMC}+p_\text{YMC})=0,\\
 &&\dot{\rho}_m+3\frac{\dot{a}}{a}\rho_m=0, \label{aa2}\\
 &&\dot{\rho}_r+3\frac{\dot{a}}{a}(\rho_r+p_r)=0. \label{aa3}
 \end{eqnarray}
From Eqs. \eqref{aa2} and \eqref{aa3}, we can immediately obtain the standard evolutions of the matter and radiation components, $\rho_m\propto a^{-3}$ and $\rho_m\propto a^{-4}$. A less obvious but still rather simple task is solving the evolution of the YMC component. Inserting \eqref{rho_p} into \eqref{aa1}, we obtain the following relation,
 \begin{eqnarray}\label{y_evolution_eq}
\dot{\theta} \left( \mathcal{W}' + 2 \mathcal{W}''\, \theta \right) + 4 \frac{\dot{a}}{a} \mathcal{W}'\, \theta =0\,,
 \end{eqnarray}
which is in a quite compact form. This equation is integrable for any regular enough $\mathcal{W}$. In particular, we can easily derive the result 
 \begin{eqnarray}\label{y_evolution}
\sqrt{\theta} \mathcal{W}'\left(\theta\right) = \alpha a^{-2}\,,
 \end{eqnarray}
where $\alpha$ is a coefficient of proportionality that depends on the initial conditions.

At very high redshift, Eq. \eqref{y_evolution} entails an increase of the order-parameter $\theta$ that involves the limit $\theta \gg \kappa^2$. Thus, at very-high redshift the system transits towards the ultraviolet regime. Within this limit, Eq. \eqref{eos} encodes the EOS parameter $w_\text{YMC}\rightarrow 1/3$. The YMC then starts behaving as the radiation component, as one would have expected since the theory is evolving towards asymptotic freedom at high energy. At small redshift, the expansion of the universe requires the LHS of Eq. \eqref{y_evolution} to vanish, which occurs for the extremal value of $\theta=\theta_0$, and the EOS' parameter then converges towards $w_\text{YMC}=-1$. This ensure that the YMC behaves as an effective cosmological constant.\\

\noindent 
Finally, notice that we may proceed as in \cite{YZ2,YZ3}, and take advantage of the observational constraint on the ratio between the dark-energy-density and the critical energy density, in order to fix the energy-scale $\kappa^2$ that appears in the effective Lagrangian $\mathcal{W}(\theta)$. The effective Lagrangian is no more dependent on the  parameter $\kappa$, and we can rescale $\theta$ as in the previous literature.

\subsection{Interacting YMC models}
\noindent 
In this section, we generalize the YMC dark energy model and take into account some effective interaction with dust matter. Nevertheless, for the sake of simplicity, at this first stage of the analysis we will disregard the interaction between radiation and the YMC. We will then describe the YMC dark energy and background matter interaction through one additional parameter $Q$. The equations of the conservation of energy in \eqref{aa1} and \eqref{aa2} should be modified into
 \begin{eqnarray}\label{b1}
&&\dot{\rho}_\text{YMC}+3\frac{\dot{a}}{a}(\rho_\text{YMC}+p_\text{YMC})=-\frac{\dot{a}}{a} Q ,\\ 
\label{b2}&&\dot{\rho}_m+3\frac{\dot{a}}{a}\rho_m=+\frac{\dot{a}}{a} Q, \\ 
&&\dot{\rho}_r+3\frac{\dot{a}}{a}(\rho_r+p_r)=0. \label{b3}
 \end{eqnarray}
The interaction parameter $Q$, in natural units, has the dimension of $[{\rm energy}]^4$ and has been introduced for phenomenological reasons. Its possible form will be addressed later. If the YMC transfers energy to the matter, for instance if the YMC decays into pairs of matter particles, we should require the parameter $Q$ to be positive. Vice-versa we should require the parameter $Q$ to be negative. \\

\noindent
We can then proceed to the study of the evolution of the system \eqref{b1}-\eqref{b2}. It is convenient to introduce the so-called e-folding time $N\equiv\ln a$, the derivative with respect to which we will denote with a prime. We will denote as $x$ the dimensionless matter density $x=\rho_m/\kappa^2$. The system \eqref{b1}-\eqref{b2} then reads 
\begin{eqnarray}
 \theta' \left(\mathcal{W}' + 2 \theta \mathcal{W}'' \right) + 4 \theta \mathcal{W}'&=&-Q,\label{y'}\\
 x'+ 3 x&=&+Q.\label{x'}
\end{eqnarray}
Using the above definitions, we can immediately recast the fractional energy densities of the YMC and the dust matter
\begin{eqnarray}
 \Omega_\text{YMC}=\frac{- \mathcal{W}+ 2 \mathcal{W}' \theta}{- \mathcal{W}+ 2 \mathcal{W}' \theta + x},~~{\rm
 and}~~\Omega_{m}=\frac{x}{- \mathcal{W}+ 2 \mathcal{W}' \theta +x}.\label{Omega}
\end{eqnarray}
\\

\noindent 
It's useful to discuss the general properties of this dynamical system before specifying the form of the interaction term $Q$. We can seek the fixed points of the system by imposing $\theta'=x'=0$ in Eqs.\eqref{y'}-\eqref{x'}, and then look for the solutions $\theta_c$ and $x_c$ of of the simplified system derived from Eqs. \eqref{y'}-\eqref{x'}. This latter reads  
\begin{eqnarray}
 4 \theta_c \mathcal{W}'(\theta_c)&=&-3 x_c,\label{yc}\\
 3 x_c&=&+Q(x_c,\theta_c).\label{xc}
\end{eqnarray}
The stability of the solutions of these differential equations, and the possible existence of attractor solutions, will be analyzed in the forthcoming subsections, in which we will specialize the form of the coupling $Q$ by assuming different linear combinations of the energy-density components considered so far. 

\subsubsection{$Q\propto \rho_\text{YMC}$}\label{case1}

\noindent
In this section, we parametrize the interaction as proportional to the YMC energy-density, namely $Q=\alpha \, \rho_\text{YMC}=\alpha \left(-\mathcal{W} + 2 \mathcal{W}' \theta \right)$. The trivial case $\alpha=0$ reduces to the free YMC dark energy model studied above. We will just consider the simplest case with $\alpha$ being a non-zero dimensionless constant. The evolution equations \eqref{y'}-\eqref{x'} then recast 
\begin{eqnarray} \label{inter11}
 \theta' \left(\mathcal{W}' + 2 \theta \mathcal{W}'' \right) + 4 \theta \mathcal{W}'&=&-\alpha \left(-\mathcal{W} + 2 \mathcal{W}' \theta \right),\\
\label{inter12} x'+ 3 x&=&+\alpha \left(-\mathcal{W} + 2 \mathcal{W}' \theta \right).
\end{eqnarray}
When the fraction\textcolor{blue}{al} density of the YMC is sub-dominant in the universe, we expect the effect on the dust to be small. Only in the latest stage of expansion of the universe, where the YMC dark energy dominates its evolution, the effect of interaction on the dust component can become important.\\

\noindent
The critical point \eqref{yc}-\eqref{xc} now rewrites 
\begin{eqnarray} \label{ccc}
 2\left(2 + \alpha \right) \theta_c \mathcal{W}'(\theta_c)&=& \alpha  \mathcal{W}(\theta_c),\\
 3 x_c&=&- 4 \theta_c \mathcal{W}'(\theta_c).
\end{eqnarray}
It's easy to see that the fractional energy densities of the YMC and the EOS at this critical point do not depend on the details of the potential $\mathcal{W}$, but might rather have a dependence on the parameter $\alpha$. It is straightforward to verify this from the very definition of EOS calculated at the fixed point solution of Eq. \eqref{ccc}:
\begin{equation}
\Omega_\text{YMC} = - \frac{1}{w_\text{YMC}}= \frac{\mathcal{W}(\theta_c)-2\theta_c\mathcal{W}'(\theta_c)}{\mathcal{W}(\theta_c)-\frac{2}{3}\theta_c\mathcal{W}'(\theta_c)} = \frac{3}{3+\alpha}\,.
\end{equation}
The constraint $0\le\Omega_\text{YMC}\le1$ implies that $\alpha>0$. Internal consistency of the model requires the solution of the system to be stable under perturbations. The next step is then to require the critical point to be an attractor solution. In order to achieve that, we first need to compute the eigenvalues of the linearized system \eqref{inter11}-\eqref{inter12} at the critical point:
\begin{eqnarray}
&\lambda_1=& -3\,, \\
&\lambda_2=& - \frac{1}{\left(\mathcal{W}'(\theta_c) + 2 \theta_c \mathcal{W}''(\theta_c) \right)^2} \left((4+\alpha)\mathcal{W}'(\theta_c)^2 + (2+\alpha)4 \theta_c^2 \mathcal{W}''(\theta_c)^2 + 4(3+\alpha) \theta_c\mathcal{W}'(\theta_c)\mathcal{W}''(\theta_c)\right) \,. 
\end{eqnarray}
The solution is an attractor if
\begin{equation}
(4+\alpha)\mathcal{W}'(\theta_c)^2 + (2+\alpha)4 \theta_c^2 \mathcal{W}''(\theta_c)^2 + 4(3+\alpha) \theta_c\mathcal{W}'(\theta_c)\mathcal{W}''(\theta_c)>0 \, .
\end{equation}
In general we will need a specific form of $\mathcal{W}$ to further discuss the nature of the critical point. 

\subsubsection{$Q\propto \rho_m$}\label{case2}
\noindent
The next case to be considered hinges on an interaction of the form $Q=\alpha \rho_m=\alpha x$. The evolution equations \eqref{y'}-\eqref{x'} now rewrites 
\begin{eqnarray}
\label{inter1}
 \theta' \left(\mathcal{W}' + 2 \theta \mathcal{W}'' \right) + 4 \theta \mathcal{W}'&=&-\alpha x,\\
 x'+ 3 x&=&+\alpha x \,.
\end{eqnarray}
If we assume $\alpha$ to be a non vanishing constant, we easily recover that $\rho_m\propto a^{\alpha-3}$. This result might then lead to observational inconsistencies: the evolution of the dust component conflicts with the evolution of dust in the standard big-bang model. We should then avoid considering such a form of interaction term in the early stage of evolution of the universe, at very high redshift.

Nevertheless, even if we insisted in phenomenologically describing at small redshift the interaction between YMC dark matter and matter energy-density with a term of the form $Q= \alpha x$, we would find only a trivial critical point $\theta_c=\theta_0$ and $x_c=0$. Thus, we must conclude that it is impossible to obtain an attractor solution for this kind of system.

\subsubsection{$Q\propto \rho_\text{YMC} + \rho_m$}\label{case3}
\noindent
As a final case, we can discuss a model where $Q=\alpha \left(\rho_\text{YMC} + \rho_m\right)$. We limit again ourselves to the consideration of $\alpha$ being a dimensionless constant. In the later stage, when the dark energy dominates the evolution, this model reduces to the case we studied in Sec.~\ref{case1}, while in the dust dominated stage reduces to the case we studied in Sec.~\ref{case2}.\\

\noindent
In analogy to the discussion developed in the previous section, if we insisted in applying this model to the description of the early universe, the evolution of dust would turn to be incompatible with the prediction of the standard hot big-bang models. Nevertheless, if we want to develop a phenomenological model for small redshift, we may elaborate on this case. 

The dynamical equations in \eqref{y'}-\eqref{x'} become
\begin{eqnarray}
\label{inter3}
 \theta' \left(\mathcal{W}' + 2 \theta \mathcal{W}'' \right) + 4 \theta \mathcal{W}'&=&-\alpha \left(-\mathcal{W} + 2 \mathcal{W}' \theta + x\right)\,,\\
 x'+ 3 x&=&+\alpha \left(-\mathcal{W} + 2 \mathcal{W}' \theta + x\right)\,,
\end{eqnarray}
and the system admits a critical point in 
\begin{equation}
3 \alpha  \mathcal{W}=2 (\alpha +6) \theta_c  \mathcal{W}', \qquad x_c=-\frac{4}{3} \theta_c \mathcal{W}'(\theta_c) \,.
\end{equation}
The fractional energy density and the EOS of the YMC at this critical point now read
 \begin{eqnarray}
\Omega_\text{YMC}=-\frac{1}{w_\text{YMC}}=\frac{3-\alpha}{3}.
\end{eqnarray}
 Notice that now the observational constraint of $0\le\Omega_\text{YMC}\le1$ sets a different range of allowed values for $\alpha$, namely $0<\alpha\leq3$.\\
 
\section{A non-perturbative example: SU(2)-YMC}
\noindent
We focus now to the analysis of the YMC model we have described in the previous sections, with a specific choice for the IR effective Lagrangian. We keep analyzing a YM theory that enjoys a SU(2) gauge group, and show at non-perturbative level that a YMC forms that drives accelerated expansion of the universe at small redshift. 
The Functional Renormalization Group (FRG) approach to non-Abelian gauge theories will be particularly fruitful to our purposes, as it allows to introduce non-perturbative methods that can be treated as much as it is possible analytically.

\subsection{Functional Renormalization Group}
\noindent 
The FRG approach is a tool developed to study interacting QFT and statistical systems in the non-perturbative regime, where no small coupling exists and perturbative techniques are not applicable. The method is based upon a Wilsonian momentum-shell wise integration of the path-integral: a mass-like regulator function $R_k(p)$ suppresses quantum fluctuations with momenta lower than an IR momentum cutoff scale $k$. This allows us to define a scale-dependent effective action, the flowing action $\Gamma_k$, which only contains the effect of quantum fluctuations with momenta greater than $k$. By changing $k$ we can interpolate smoothly  between the microscopic action $\Gamma_{k \rightarrow \infty}$ and the full quantum effective action $\Gamma_{k \rightarrow 0}$.
The scale-dependence of the flowing action is then given by the Functional Renormalization Group Equation (FRGE) \cite{Wetterich:1993yh} and \cite{Morris:1993qb}:
\begin{equation}
\label{WetterichEq}
\partial_{t}\Gamma_{k}=\frac{1}{2}\mathrm{STr}\left(\Gamma_{k}^{(2)}+R_{k}\right)^{-1}\partial_{t}R_{k}\,.
\end{equation}
Herein, $\Gamma_k^{(2)}$ denotes the second functional derivative of the flowing action with respect to the fields, and constitutes a matrix in field space. The Super-Trace $\rm STr$ includes a summation over all discrete indices and the fields, including a negative sign for Grassmann valued fields, {\it i.e.} fermions and Faddeev-Popov ghosts. The Super-Trace also includes a summation over the eigenvalues of the Laplacian in the kinetic term. The main technical advantage of the FRGE lies in its one-loop form, which nevertheless takes into account higher-loop effects, as it depends on the full, field-dependent nonperturbative regularized propagator $\left(\Gamma_k^{(2)}+R_k \right)^{-1}$ (see \cite{Papenbrock:1994kf}).
The FRGE has been extensively applied to $SU(N)$ Yang-Mills theories, for further references see \cite{Fischer:2002hn,Fischer:2006vf,Fischer:2008uz,FRG,Pawlowski:2003hq} and for the application of the FRG to the study of YMC \cite{Reuter:1994zn,Reuter:1997gx,Gies:2002af,Eichhorn:2010zc}.

\subsection{Finding the effective Lagrangian}

\noindent
Solving the equation \eqref{WetterichEq} exactly is a titanic task. Since we are mainly interested in qualitative and as much as possible analytic results, we will deploy some approximations. (For the reader  interested in the state of art of YMC in the FRG framework, we refer to the work by Eichhorn, Gies and Pawlowski \cite{Eichhorn:2010zc}, in which a numerical extrapolation between low and high momenta of full propagators were used to compute the gluon condensate.)

First of all we will replace $\Gamma_{k}$ in the RHS of equation \eqref{WetterichEq} with the bare action $S$\footnote{This kind of approximation is usually called ``one loop'' in the FRG literature because of the similarities between the FRGE and the standard one-loop effective action.}, doing so we are allowed to integrate both sides of the equation
\begin{equation*}
\Gamma_k=-\int \mathcal{L}_{\rm eff}=-\int \mathcal{W}_{k}(\theta)=\int dk \frac{1}{2}\mathrm{STr}\left(S^{(2)}+R_{k}\right)^{-1}\partial_{t}R_{k}\, = \frac{1}{2}\mathrm{STr}\,\mathrm{Log}\left(S^{(2)}+R_{k}\right) + const.
\end{equation*}
We select the bare action to be $S=\frac{1}{4}\int\mathrm{d}x\, F_{a}^{\mu\nu}F_{\mu\nu}^{a}$, that corresponds to the UV limit of our effective theory. We will fix the integration constant requiring the effective action to vanish for vanishing field strength.

To correctly invert the regularized propagator we need to include in the action a (harmonic) gauge-fixing and the associated Faddeev-Popov ghosts.
\begin{equation*}
S_{\rm gf}=\frac{1}{2\alpha}\int\mathrm{d}x\,\overline{D}_{\mu}a_{\nu}^{a}\overline{D}_{\nu}a_{\mu}^{a}\,,\qquad S_{\rm gh}=\int\mathrm{d}x\,\overline{D}_{\mu}\bar{c}_{\nu}D^{\mu}c^{\nu}\,.
\end{equation*}
In the Landau gauge $\alpha\to0$, the trace over the gauge field space is restricted to the transverse sector
\begin{equation}
\label{tracce}
\frac{1}{2}\mathrm{STr}\,\mathrm{Log}\left(S^{(2)}+R_{k}\right)  =\frac{1}{2}\mathrm{Tr_{transv.}}\,\mathrm{Log}\left(\bar{D}_{T}^{\mu\nu}+R_{k}\left(\bar{D}_{T}^{\mu\nu}\right)\right)-\frac{1}{2}\mathrm{Tr_{ghost}}\,\mathrm{Log}\left(\bar{D}_{gh}^{\mu\nu}+R_{k}\left(\bar{D}_{gh}^{\mu\nu}\right)\right)\,,
\end{equation}
with operators $\bar{D}_{T}^{\mu\nu}=\overline{\square}\delta_{cb}\delta^{\mu\nu}+g\overline{F}^{a\mu\nu}f_{abc}$ and $\bar{D}_{gh}^{\mu\nu}= \eta^{\mu\nu} \overline{\square}$, in which $g$ is the $YM$ coupling and the barred quantities are made of background fields; for more details on the actions and its variations we refer to the Appendix \ref{actionYM}.
We will employ the simplest possible regulator functions (mass-like cutoff)
\begin{equation}
\label{regulator}
R_{k}\left(\mathcal{D}\right)=k^2 ,
\end{equation}
in both the transversal gauge and ghost sector $\mathcal{D}\to\bar{D}_{T}^{\mu\nu},\,\bar{D}_{gh}^{\mu\nu}$.\\
We can employ an integral representation\footnote{One should actually be more careful with the definition of the integral. A more precise formula is the following: $$\mathrm{Log}\left(A\right) = - \lim_{\epsilon\to 0}\int_\epsilon^\infty \frac{ds}{s} \left(e^{-A s}-e^{-s}\right).$$ Nevertheless we will use the naive representation and implicitly regularize the final expression.} of the logarithm, in order to find an exactly summed expression
\begin{equation}
\mathrm{Log}\left(A\right) = - \int_0^\infty \frac{ds}{s} e^{-A s}\,.
\end{equation}
A wise choice of the background allow us to perform the traces as sums over the eigenvalues of the chosen kinetic operators. In general the eigenvalues of the operator $\bar{D}_{T}$ are not known, and the only known stable covariantly constant background is the self-dual one already employed in the FRG context in \cite{Eichhorn:2010zc} (the key properties needed for this work are summarized in the Appendix \ref{selfdual}). The effective Lagrangian is finally recovered to be

\begin{align} \label{adricat}
\mathcal{L}_{\rm eff}  =& \,\frac{g^2 B^2}{2 \pi^2} \int_0^\infty \frac{ds}{s} \sum_{m,n=0}^\infty \left(e^{-2gB\left(n+m\right) + k^2}+e^{-2gB\left(n+m+2\right) + k^2}-e^{-2gB\left(n+m+1\right) + k^2}\right) \nonumber\\
&\!\!\!\!\!\!= \frac{g^2 B^2}{2 \pi^2} \int_0^\infty \frac{ds}{s}e^{-k^2 s}\left(\frac{1}{4\sinh^2\left(g B s\right)}+1\right)\,.
\end{align}
The ``magnetic field'' $B$ is the only variable of the self-dual background, which is related to the order parameter through $\theta=B^2$. The next step is to remove the constant part from the integral that gets contribution from the lowest order expansion of the $\sinh$. We then perform a change of variable, and reshuffle \eqref{adricat} as
\begin{align} \label{eqLeff}
\mathcal{L}_{\rm eff} = \mathcal{W}_{k}(\theta) = \frac{g^2 B^2}{2 \pi^2} \int_0^\infty \frac{ds}{s}e^{-\frac{k^2}{g B} s}\left(\frac{1}{4\sinh^2\left(s\right)}+1 - \frac{1}{4s^2}\right)=\frac{g^2 \theta}{2 \pi^2} \int_0^\infty \frac{ds}{s}e^{-s\left(\frac{k^4}{g^2 \theta}\right)^{1/2}}\left(\frac{1}{4\sinh^2\left(s\right)}+1 - \frac{1}{4s^2}\right)\,.
\end{align}

The asymptotic behavior for small coupling constant $g$ of the integral is reproduced exactly at the lowest order the one-loop Effective action. Furthermore, the (unique) non trivial minimum for \eqref{eqLeff} is found to be at $\frac{g^2\,\theta_0}{k^{4}} \approx 0.361$, as it can be read out from Fig.1. 
\begin{figure}[!h]
\label{picLeff}
\includegraphics[scale=1]{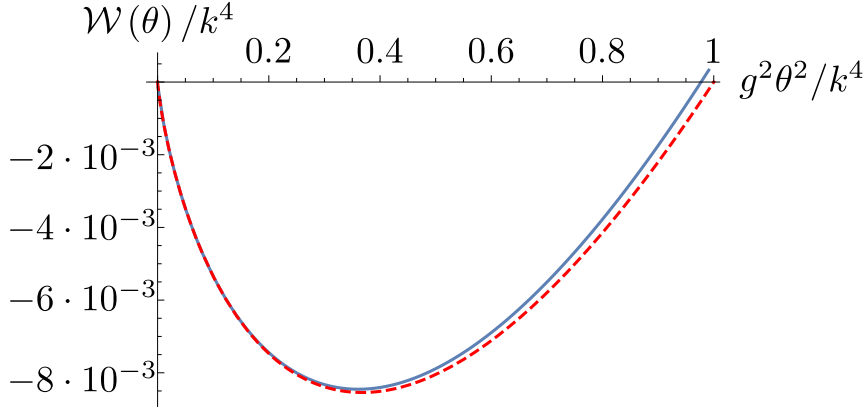}
\caption{Plot of the function \eqref{eqLeff} (blue-continuous) and the one-loop \eqref{l_eff} (red-dashed). Notice the presence of a non zero global minimum for $\frac{g^2 \theta_0}{k^{4}} \approx 0.361$.}
\end{figure}
\subsection{FRG improved YMC Lagrangian}
\noindent
In order to check whether the effective Lagrangian calculated in \eqref{eqLeff} and derived for a SU(2) YM theory can actually explain dark energy, we need to review if our example satisfies the properties we discussed in Section \ref{snp}.
\begin{enumerate}
\item
From Fig.1 
it's evident that the function \eqref{eqLeff} has a non zero global minimum. The exact position of this minimum can be computed numerically, and in terms of dimensionless quantities is found to be $\frac{g^2\,\theta_0}{k^{4}} \approx 0.361$.
\item It is possible to compute the asymptotic expansion of $ \mathcal{W}_{k}(\theta)$ for small values of the YM coupling constant $g$. In this limit we are able to reproduce the one-loop result derived by Savvidy \cite{Savvidy:1977as}: if we perform the Taylor expansion of $1/\sinh^2(s) =\frac{1}{s^2}-\frac{1}{3}+O\left(s^2\right)$, we find indeed
\begin{equation}
 \mathcal{W}_{k}(\theta)  = \frac{g^2 \theta}{2 \pi^2} \int_0^\infty \frac{ds}{s}e^{-s\left(\frac{k^4}{g^2 \theta}\right)^{1/2}}\left(-\frac{1}{12}+1\right) + \ldots =  \frac{11}{24 \pi^2} g^2 \theta \int_0^\infty \frac{ds}{s}e^{-s\left(\frac{k^4}{g^2 \theta}\right)^{1/2}}+ \ldots = \frac{1}{2} \frac{11}{24 \pi^2} g^2 \theta \mathrm{Log} \left(\frac{k^4}{g^2 \theta}\right) \,.
\end{equation}
\item $ \mathcal{W}_{k}(\theta)$ shows an asymptotic behavior at least linear in $\theta$. This means that for $\theta\gg k^4$ the exponential in the integral tends to $1$ and the only $\theta$ dependence is the overall one, namely $$\mathrm{Log}\left( \mathcal{W}_{k}(\theta) \right)=\mathrm{Log}\left(\theta\right)+ O(\theta).$$  The condition $
\frac{d \ln \mathcal{W}}{\!\!d \ln \theta} \to 1$ then follows immediately.         
\end{enumerate}
\noindent
In the following sections, we show in detail that the YMC described by our toy model evolves from a radiation like component to a dark energy one.
\subsubsection{A non-interacting YMC model}
\noindent 
We have already shown that in the case of a non-interacting YMC model the condensate evolution equation is implicitly solvable, and that the solution $\theta(a)$ can be obtained by inverting the equation
 \begin{eqnarray}
\sqrt{\theta} \mathcal{W}_k'\left(\theta\right) = \alpha a^{-2}\,,
 \end{eqnarray}
where $\alpha$ is a coefficient of proportionality that depends on the initial conditions. We can then fix the renormalization scale $k$ by comparing the ``predicted'' YMC fractional energy density with the measured Dark Energy fractional energy density ($\Omega_\Lambda = 0.735$) finding for a big range of initial conditions $k\approx 3.2 h^{1/2} 10^{-3} eV$. This energy scale, as already noted in \cite{YZ1,YZ2,YZ3}, is low compared to typical energy scales in particle physics, such as the QCD and the weak-electromagnetic unification, and this prevents the identification of this YMC as a condensate of some SM gauge fields. 
Then we can study the evolution of the YMC energy density and its EOS for different values of $\alpha=(10^{-7},10^{-5})$, still obtaining the same asymptotic values. Results are summarized in Fig. 2.
\begin{figure}[!h]
\label{pic2}
\includegraphics[scale=.85]{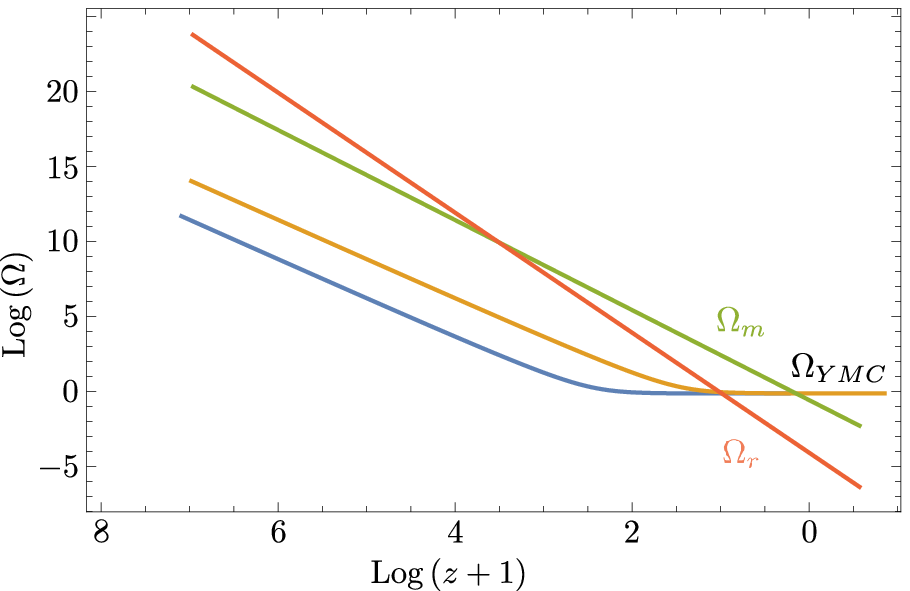}\hspace{0.5cm}\includegraphics[scale=.88]{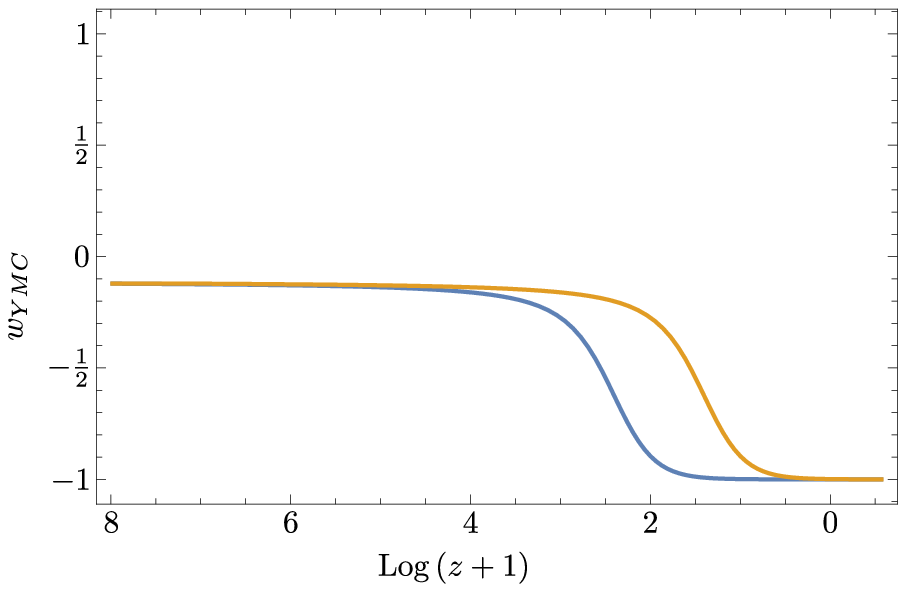}
\caption{In the free YMC dark energy model, the evolution of the YMC fractional energy density (left panels) and EOS (right panels) for the models with different initial conditions ($\alpha=10^{-7}$ on the left and $\alpha=10^{-5}$ on the right).}
\end{figure}
\subsubsection{$Q\propto \rho_\text{YMC}$}
\noindent
For YMC models enjoying an interaction proportional to the YMC energy density, we have already discussed in Section \ref{case1} the equation defining the existence of a fixed point --- derived from the differential equation evolution system  \eqref{y'}-\eqref{x'} --- and the condition to be imposed on the coupling parameter $\alpha$ in order to characterize an attractor solution. Here we report numerical computations on the position of the fixed point and on the value of the critical exponents at the fixed point. The fixed point exists for every positive value of the coupling parameter and is always attractive, as shown in Fig. 3. 
\begin{figure}[!h]
\label{pic3}
\includegraphics[scale=0.85]{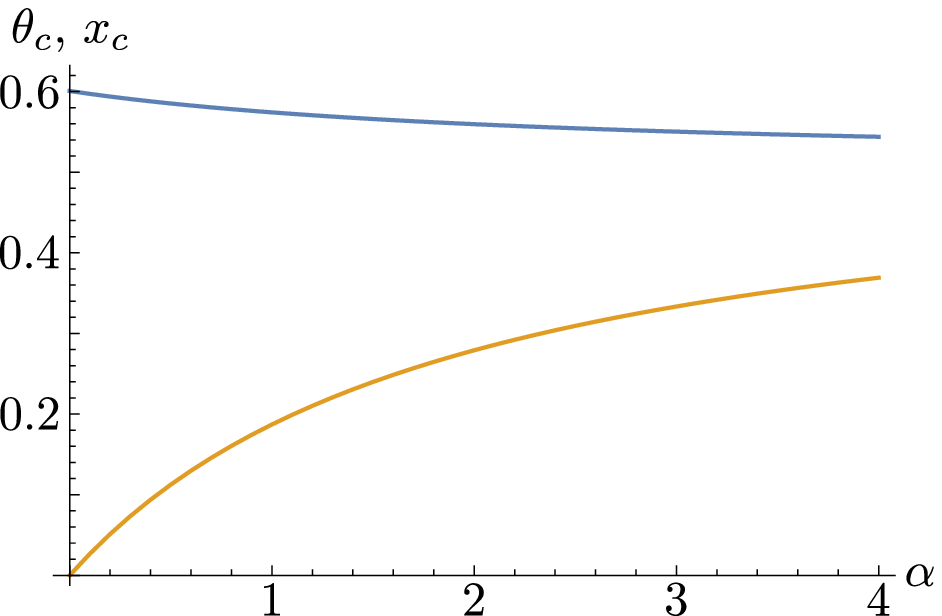}\hspace{.5cm}
\includegraphics[scale=0.85]{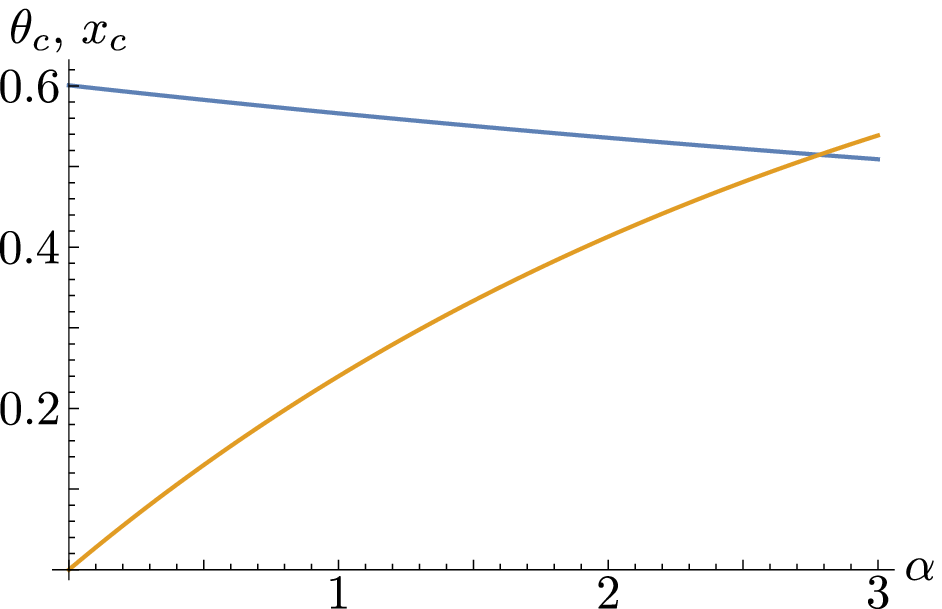}\hspace{.5cm}\includegraphics[scale=0.85]{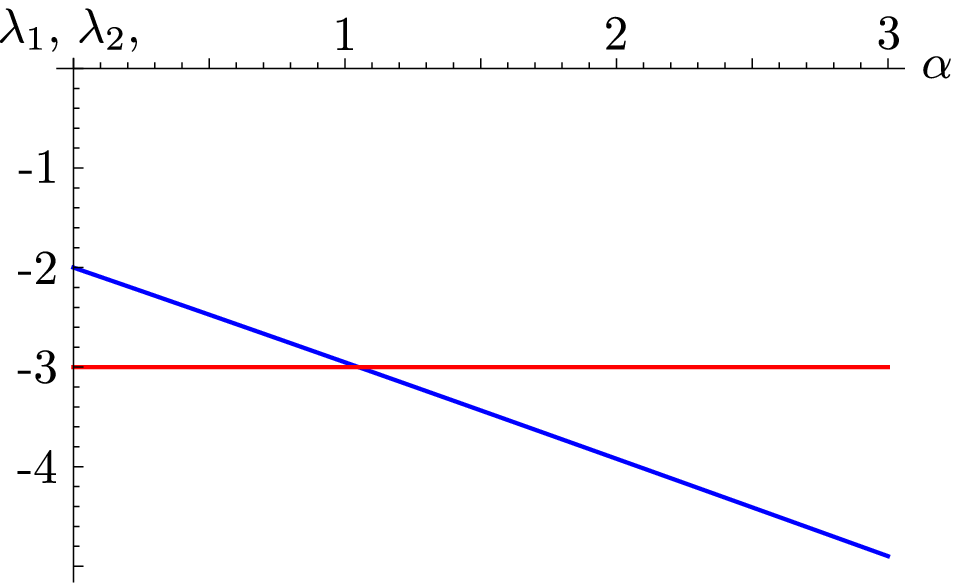} \hspace{.5cm}
\includegraphics[scale=0.85]{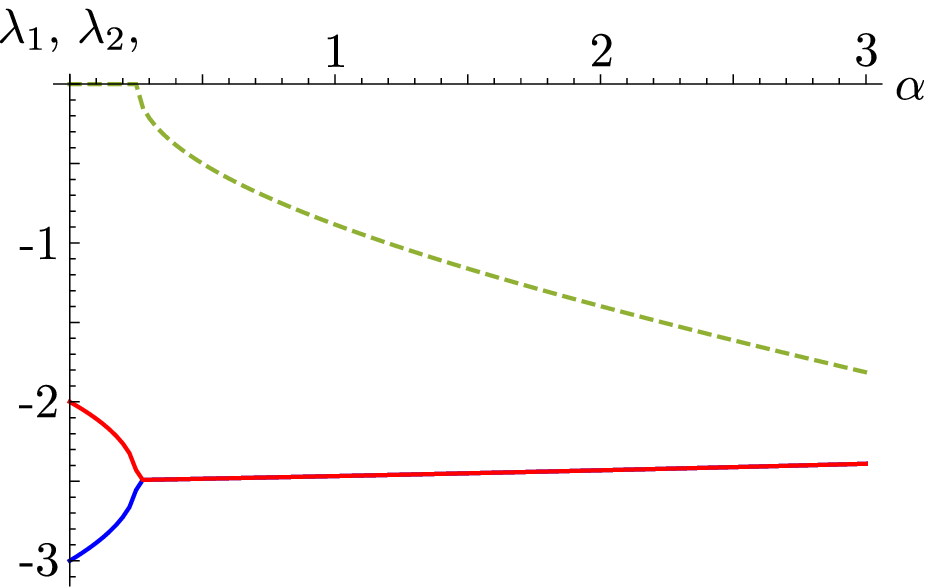}
\caption{For the coupled YMC dark energy models, the plot against the coupling parameter $\alpha$ of the critical order parameter and matter density $\theta_c, x_c$ (top panels) and of the critical exponents $\lambda_1,\lambda_2$ at the fixed point (bottom panels). In the bottom panel, on the left is reported the case in which $Q\propto \rho_\text{YMC}$, and on the right the case in which $Q\propto \rho_\text{YMC} + \rho_\text{m}$.}
\end{figure}
\section{Concluding remarks}
\noindent 
The query whether the YMC may actually provide a consistent and physically reliable model for dark energy, along the lines of the analysis first developed in Refs~\cite{YZ1,YZ2,YZ3}, has been addressed in this paper within the attempt of finding support to this theoretical hypothesis in the non-perturbative approach to the calculation of the Yang-Mills effective action. 

In particular, we have first discussed the properties that the effective Lagrangian $\mathcal{W}(\theta)$ must posses in order to drive the Universe towards a cosmological dark energy phase. If a condensate exists, {\it i.e.} if the non-perturbative effective action has a minimum in the YMC order-parameter $\theta$, then the model can actually reproduce the dark energy behavior of the expanding universe at small redshift. If the effective action scales at least like the bare Yang-Mills action for high-energy scale, at high redshifts it entails the EOS of radiation. Moreover, internal consistency also requires that perturbative one-loop results must be still recovered in the appropriate asymptotic limit.

We have then focused on the particular example provided of the SU$(2)$ Yang-Mills bare action. We have deployed non-perturbative techniques mutated from the FRG method, in order to show that for the SU$(2)$ Yang-Mills a non-trivial minimum indeed exists, and that the high-energy scale regime approaches known perturbative results, and yields radiation dominated EOS. Despite our conclusions can be only based so far on this particular example, this successful check of the  requirements necessary to have a viable YMC dark energy model seems to us extremely encouraging in strengthening the possibility that YMC can work as a model for dark energy. 

The improvement of the non-perturbative techniques may allow in the future to extend the present analysis to the cases of SU$(N)$ gauge-groups, or more in general to other classes of Lie-groups. At this purpose, we may either ask ourselves whether condensation can work for any SU$(N)$ group, or whether consistency of the model necessarily predicts a maximal values of $N$ in order the mechanism to work. Not disconnected to these questions, it comes the query on the relation between the Yang-Mills fields involved, which are necessary in order the condensate to form, and the elementary-particle fields advocated to explain dark matter. Indeed, it would be tempting to try to link YMC dark energy models to other models for dark matter, such as the ones referred to in the literature \cite{M1, M2, M3, M4, M5, M6} as Mirror Standard Model for dark matter.

\section*{Acknowledgment}
\noindent
We are grateful to Yifu Cai and Roberto Percacci for extremely useful discussions.

\appendix

\section{Action and variations}\label{actionYM}

\noindent
We consider the bare action of the Yang-Mills theory 

\begin{equation*}
\mathcal{S}=\frac{1}{4}\int\mathrm{d}x\, F_{a}^{\mu\nu}F_{\mu\nu}^{a}\,.
\end{equation*}

We split the field $A_{\mu}$ into a background plus a fluctuation
$A_{\mu}=\bar{A}_{\mu}+a_{\mu}$. The quadratic part of this action
in the fluctuation field is 

\begin{align*}
\delta^{2}S & =\frac{1}{2}\int\mathrm{d}x\,\left(\overline{D}_{\mu}a_{\nu}^{a}\overline{D}_{\mu}a_{\nu}^{a}-\overline{D}_{\mu}a_{\nu}^{a}\overline{D}_{\nu}a_{\mu}^{a}+g\overline{F}^{a\mu\nu}f_{abc}a_{\nu}^{b}a_{\mu}^{c}\right)\\
 & =\frac{1}{2}\int\mathrm{d}x\, a_{\nu}^{b}\left(\delta_{cb}\overline{\square}+\overline{D}^{\mu}\overline{D}^{\nu}\delta_{cb}+g\overline{F}^{a\mu\nu}f_{abc}\right)a_{\mu}^{c}\,,
\end{align*}
where the bar quantities are made out of the background fields. To compute the inverse propagator we need to add also a gauge fixing action and the corresponding Faddeev-Popov ghost action:
\begin{equation*}
S_{gf}=\frac{1}{2\alpha}\int\mathrm{d}x\,\overline{D}_{\mu}a_{\nu}^{a}\overline{D}_{\nu}a_{\mu}^{a}\,,\qquad S_{gh}=\int\mathrm{d}x\,\overline{D}_{\mu}\bar{c}_{\nu}D^{\mu}c^{\nu}
\end{equation*}
The FRGE splits into the trace over the transverse part and the longitudinal
part of the connection field and the ghost sector:
\begin{align*}
\frac{1}{2}\mathrm{STr}\left(S^{(2)}+R_{k}\right)^{-1}\partial_{t}R_{k} & =\frac{1}{2}\mathrm{Tr}_{T}\left(S^{(2)}+R_{k}\right)^{-1}\partial_{t}R_{k}+\frac{1}{2}\mathrm{Tr}_{L}\left(S^{(2)}+R_{k}\right)^{-1}\partial_{t}R_{k}-\mathrm{Tr}_{gh}\left(S_{gh}^{(2)}+R_{k}\right)^{-1}\partial_{t}R_{k}\\
 & =\frac{1}{2}\mathrm{Tr}\frac{\partial_{t}R_{k}}{\bar{D}_{T}^{\mu\nu}+R_{k}}+\frac{1}{2}\mathrm{Tr}\frac{\alpha\partial_{t}R_{k}}{\bar{D}_{L}^{\mu\nu}+\alpha R_{k}}-\mathrm{Tr_{ghost}}\frac{\partial_{t}R_{k}}{\square+R_{k}}\,.
\end{align*}
Calculations are simplified by the fact that in the Landau gauge $\alpha\to0$ the longitudinal trace can be dropped out completely.
\section{Self-dual field configuration}\label{selfdual}
\noindent 
We may choose a background field configuration that allows to project
onto the effective potential $\mathcal{W}(\theta)$. Hence a covariantly constant
field strength with $D_{\mu}F^{\mu\nu}=0$ suffices. Since the spectrum
of the Laplace-type operators, like $D_{T}^{\mu\nu}=\overline{\square}\delta_{cb}\delta^{\mu\nu}+g\overline{F}^{a\mu\nu}f_{abc}$,
or at least the heat-kernel trace for these operators has to be known,
we have a limited choice in the possible background field configurations.
To avoid problems with tachyonic modes, which indicate the instability
of a background, we project onto the only known stable covariantly
constant background, which is self-dual, namely $\tilde{F}_{\mu\nu}=\frac{1}{2}\epsilon_{\mu\nu}^{\phantom{\mu\nu}\rho\sigma}F_{\rho\sigma}=F_{\mu\nu}$.
Then we set $F_{\mu\nu}=0$. Apart from $F_{01}=F_{23}\equiv B=const.$
all other non-zero components follow from the antisymmetry of the
field strength tensor. Due to the enhanced symmetry properties connected
to the self-duality, zero modes exist. These carry important information
and have to be regularized carefully, since the standard choice for
$R_{k}$ is zero on the zero mode subspace.

\begin{align*}
\mathrm{spec}\left(\bar{D}_{T}^{\mu\nu}\right) & =2gB_{l}\left(n+m+2\right)\text{ with }n,\, m\in\mathbb{N}\text{ and with multiplicity 2 in 4 dimensions,}\\
 & =2gB_{l}\left(n+m\right)\text{ with }n,\, m\in\mathbb{N}\text{ and with multiplicity 2 in 4 dimensions.}\\
\mathrm{spec}\left(\square\right) & =2gB_{l}\left(n+m+1\right)\text{ with }n,\, m\in\mathbb{N}
\end{align*}

with a degeneracy factor $\frac{B^{2}}{2\pi^{2}}$. Herein
$B_{l}=\left|\nu_{l}\right|B$ and $\nu_{l}$ is given by $\nu_{l}=spec\left\{ \left(T^{a}n^{a}\right)^{bc}\ |\ n^{2}=1\right\} $
with the generators of the adjoint representation $T^{a}$ and therefore
for a general gauge group depends on the direction of the unit vector
$n$.

\end{document}